\documentclass[aps,prb,twocolumn,groupedaddress]{revtex4}

\usepackage{graphicx}
\usepackage{dcolumn}
\usepackage{bm}

\begin{document}


\title{High Field ESR Study of the $\pi$-\textit{d} Correlated Organic Conductor \\ $\lambda$-(BETS)$_2$Fe$_{0.6}$Ga$_{0.4}$Cl$_4$}


\author{Y. Oshima}
\email[]{oshima@magnet.fsu.edu}

\author{E. Jobiliong}
\author{T. Tokumoto}
\author{J. S. Brooks}
\author{S. A. Zvyagin}
\author{J. Krzystek}
\affiliation{National High Magnetic Field Laboratory, Florida State University, Tallahassee FL 32310, U.S.A.}

\author{H. Tanaka}
\affiliation{National Institute of Advanced Industrial Science and Technology, Tsukuba, Ibaraki 305-8568, Japan}

\author{A. Kobayashi}
\affiliation{Research Centre for Spectrochemistry, Graduate School of Science, The University of Tokyo, Bukyo-ku, Tokyo 113-0033, Japan}

\author{H. Cui}
\author{H. Kobayashi}
\affiliation{Institute for Molecular Science and CREST, JST, Myodaiji, Okazaki, 444-8585, Japan}


\date{\today}

\begin{abstract}
Submillimeter and millimeter wave electron spin resonance
(ESR) measurements of the $\pi$-\textit{d} correlated organic conductor 
$\lambda$-(BETS)$_2$Fe$_{0.6}$Ga$_{0.4}$Cl$_4$ have been
performed. Antiferromagnetic resonance (AFMR) has been observed in
the insulating  antiferromagnetic phase, and its frequency-field
dependence can be reproduced by the biaxial anisotropic AFMR
theory. We find that in this alloy system, the easy-axis is near
the \textit{b}-axis, unlike previous results for the pure
$\lambda$-(BETS)$_2$FeCl$_4$ salts where it is closer to the
\textit{c}$^*$-axis. We have also observed electron paramagnetic resonance
(EPR) in the metallic phase at higher fields where the \textit{g}-value is
shown to be temperature and frequency dependent for field applied
along the \textit{c}$^*$-axis. This behavior indicates the
existence of strong $\pi$-\textit{d} interaction. Our measurements
further show the magnetic anisotropy associated with the anions
(the \textit{D} term in the spin Hamiltonian) is $|D|\sim$0.11
cm$^{-1}$.
\end{abstract}


\maketitle


\section{Introduction}
The discovery of the field-induced superconductivity in
$\lambda$-(BETS)$_2$FeCl$_4$, where BETS is
bis(ethylenedithio)tetraselenafulvalene, by Uji \textit{et al}.
has attracted considerable interest since the application of a
sufficiently strong magnetic field usually destroys the
superconducting state.\cite{ujinature} The field-induced
superconducting (FISC) phase of the FeCl$_4$ salt appears above 17
T at a temperature of 0.1 K when the magnetic field is applied parallel
to the \textit{c}$^*$-axis. This FISC phase can be explained by
the Jaccarino-Peter compensation effect where the internal
magnetic field created by the Fe$^{3+}$ moments through the
exchange interaction is compensated by the external magnetic
field, and the Zeeman effect that normally destroys the
superconductivity is suppressed under this condition.
\cite{j-peffect, alloy-uji, luisprl} In addition, the FeCl$_4$ salt shows
a metal-insulator transition at $T_{MI}$=8.3 K for zero magnetic
field that is associated with the antiferromagnetic ordering of
the Fe$^{3+}$ moments.\cite{brossard} Therefore, the magnetic
interaction between the $\pi$- and \textit{d}-electrons plays an
important role in these phenomena, and several electron spin
resonance (ESR) measurements have been performed to probe its
nature.\cite{brossard, toyota, isaak} Electron paramagnetic
resonance (EPR) and antiferromagnetic resonance (AFMR) are
observed in the paramagnetic metal (PM) and antiferromagnetic
insulating (AFI) phase, respectively.\cite{brossard, toyota,isaak} 
Resonant features, which are related to the depolarization
regime, are also observed in the canted AFI phase.\cite{isaak} 
However, no ESR measurements (i.e. resonance experiments) have
been performed in the high magnetic field region to investigate the
electronic ground state in the high field metallic and FISC
phases. Therefore, the purpose of the present work is to
explore the electronic state of the system by covering all 
aspects of the phase diagram. Recent studies have shown that the
FISC phase of organic alloys
$\lambda$-(BETS)$_2$Fe$_x$Ga$_{1-x}$Cl$_4$ shifts to lower field
as \textit{x} decreases.\cite{alloy-uji} Thus, we have focussed
on $\lambda$-(BETS)$_2$Fe$_{0.6}$Ga$_{0.4}$Cl$_4$ where the
antiferromagnetic insulating (AFI) phase exists below ~6 K and ~8 T,
and the FISC exists above ~10 T (Fig. 1), within the limits of our ESR
instrumentation.

\begin{figure}
\includegraphics[width=0.95\linewidth]{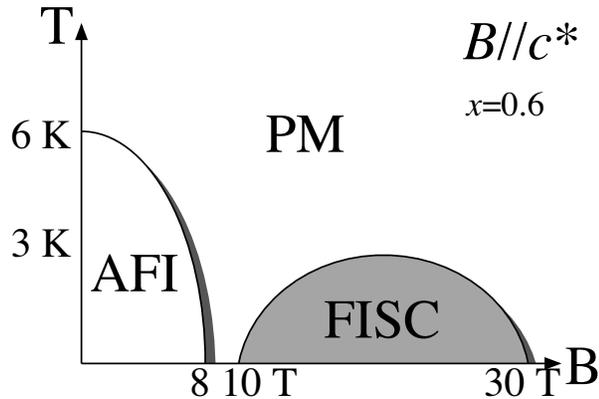}%
\caption{\label{fig:one} The phase diagram of
$\lambda$-(BETS)$_2$Fe$_{x}$Ga$_{1-x}$Cl$_4$ (\textit{x}=0.6) for
\textit{B}//\textit{c}$^*$-axis.\cite{alloy-uji} PM, AFI and FISC denote
paramagnetic metal, antiferromagnetic insulator and field-induced 
superconductor, respectively.}
\end{figure}

\section{Experimental}

The crystal structure of the series of
$\lambda$-(BETS)$_2$Fe$_x$Ga$_{1-x}$Cl$_4$ alloys has triclinic
symmetry.\cite{crystal} The planar BETS molecules are stacked
along the \textit{a}-axis and have also intermolecular
interactions along the \textit{c}-axis, which form a 2D electronic
structure. The insulating Fe$_x$Ga$_{1-x}$Cl$_4$ layer is
intercalated between these BETS layers. A finite exchange
interaction between the $\pi$-electrons of BETS molecules and the
Fe$^{3+}$ 3\textit{d} electrons (\textit{S}=5/2) is expected due
to the short inter-atomic distance between them. The samples are
needle-shaped where the needle axis corresponds to the
\textit{c}$^*$-axis.

ESR measurements were performed by using two different kinds of
techniques. The cavity perturbation technique and the single-pass transmission
technique have been used for the investigation of the low-field
region (i.e. PM and AFI phase) and the high-field region (i.e. PM
and FISC phase), respectively. The combination of an 8 T
superconducting magnet and millimeter-wave vector network analyzer
(MVNA) has been used for cavity perturbation technique.\cite{mvna} 
The MVNA includes a tunable microwave source that covers
the frequency range of 8-350 GHz, and a highly sensitive detector.
The sample was set on the end-plate of the cavity so that the
oscillatory magnetic field is always applied to the sample. We
note that this is the usual configuration for ESR measurements. 
For work at higher fields and frequencies, the 25T resistive magnet and the backward wave
oscillator (BWO) light source have been used with the
transmission technique. The BWO can cover the frequency range from
200 to 700 GHz by using several vacuum tubes. The sample was
placed in the Voigt configuration (i.e. the \textit{dc} magnetic
field is perpendicular to the propagation of the light) and a
metallic foil was placed around the sample to mask the background
radiation. The details of this high-field millimeter and
submillimeter wave facility can be found elsewhere.\cite{sergei}

\section{Results}
\subsection{AFI and PM phase below 8 T}

\begin{figure}
\includegraphics[width=0.95\linewidth]{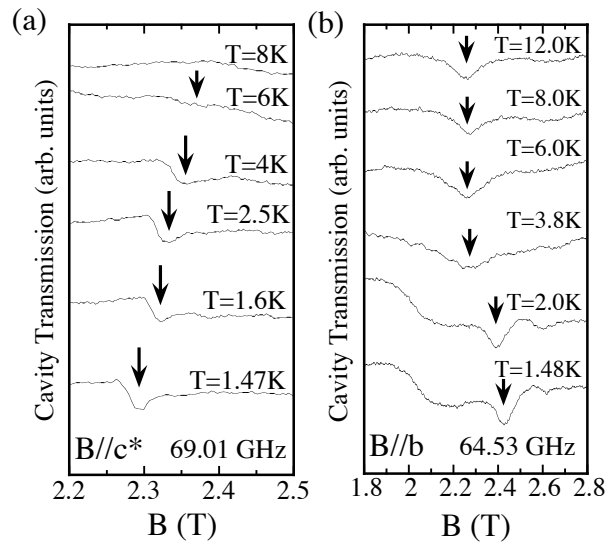}%
\caption{\label{fig:two} The temperature dependence of the ESR
spectra of $\lambda$-(BETS)$_2$Fe$_{0.6}$Ga$_{0.4}$Cl$_4$ for
(a) \textit{B}//\textit{c}$^*$-axis at 69.01 GHz, and (b) \textit{B}//\textit{b}-axis at 64.53 GHz.}
\end{figure}

Figure \ref{fig:two} (a) shows the temperature dependence of a typical
spectra where the magnetic field is applied parallel to the
\textit{c}$^*$-axis. 
The cavity perturbation technique was employed for the investigation 
of this low-field region. 
A single absorption line appears below the antiferromagnetic transition
temperature $T_{AFI}$=6 K, and the resonance field shifts to
lower field as the temperature is decreased. This shift indicates
the evolution of the internal field 
below the antiferromagnetic transition. In contrast, Fig.
\ref{fig:two} (b) shows the temperature dependence of a typical
spectra where the magnetic field is applied parallel to the
\textit{b}-axis. Unlike the \textit{c}$^*$-axis results, the
resonance shifts to higher field with decreasing temperature below
$T_{AFI}$ which is the typical behavior of AFMR in systems with a biaxial anisotropy. 
The EPR above 6 K is clearly seen up to 50 K for the \textit{b}-axis orientation
(partially shown in Fig. \ref{fig:two} (b)). 
Although the resonance field is almost independent of the temperature,
the amplitude of this absorption line decreases with the increased temperature. 
The EPR for \textit{B}//\textit{c}$^*$ was not observed
due to the weak intensity (expected for this orientation).

\begin{figure}
\includegraphics[width=0.95\linewidth]{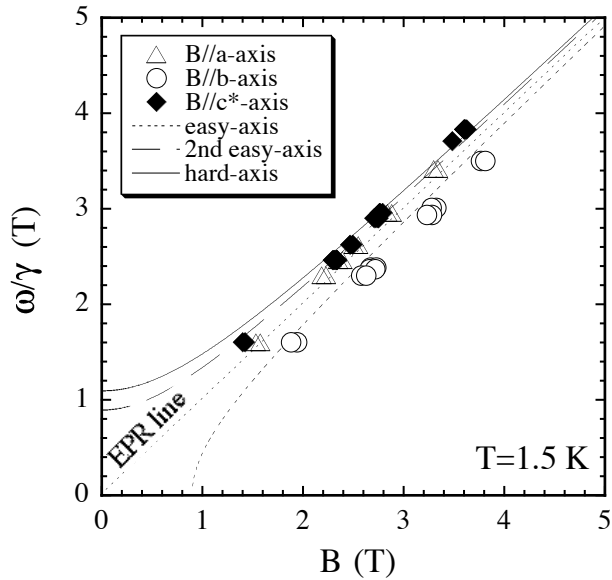}%
\caption{\label{fig:four} The frequency-field diagram of the observed AFMR at 1.5 K fitted
with the theoretical AFMR curves. (See text for details.)}
\end{figure}

Figure \ref{fig:four} shows the frequency-field diagram fit with
conventional AFMR theory.\cite{afmr} The frequency
is renormalized by the angular frequency $\omega$ divided by the
gyromagnetic ratio $\gamma$ using \textit{g}=2. 
The resonance plots are represented by triangles, circles and diamonds
 for \textit{B}//\textit{a}, \textit{b} and
\textit{c}$^*$-axes, respectively. The dotted, broken and solid
lines are the theoretical curve for the easy, 2$^{nd}$ easy and
hard axes, respectively, where the resonance conditions are as
follows:
\begin{eqnarray}
\left(
\frac{\omega}{\gamma}
\right)^2_{\mathit{easy}}= B^2-C_1,
\label{eq:AFMR1}
\end{eqnarray}
\begin{eqnarray}
\left(
\frac{\omega}{\gamma}
\right)^2_{\mathit{2^{nd}} \mathit{easy}}= B^2+C_1,
\label{eq:AFMR2}
\end{eqnarray}
\begin{eqnarray}
\left(
\frac{\omega}{\gamma}
\right)^2_{\mathit{hard}}= B^2+C_2.
\label{eq:AFMR3}
\end{eqnarray}
Here, $C_1$ and $C_2$ are the parameters $\sqrt{C_{i}}=\sqrt{2 H_E
H^i_A}$ (\textit{i}=1,2), and $H_E$, $H^1_A$, and $H^2_A$
represent the exchange field, and the anisotropy fields for the
intermediate and  hard axis, respectively. The spin-flop field
$B_{sf}$, equal to $\sqrt{C_1}$, can be obtained from the fitted
parameters. The resonance plots are best fit with $\sqrt{C_1}$=
$B_{sf}$ =0.9 T and $\sqrt{C_2}$=1.1 T for 1.5 K, where the
difference between these two values indicates that the spin system
takes a biaxial anisotropy which is consistent with similar
findings for the pure FeCl$_4$ salt.\cite{toyota} However, our
results show that the easy axis is close to the \textit{b}-axis,
contrary to the pure FeCl$_4$ salt results where the easy axis is near
the \textit{c}$^*$-axis.\cite{brossard, toyota, isaak} 
We note that we could not observe AFMR near the spin-flop field
$B_{sf}$ with X-band ESR measurements.

\subsection{FISC and PM phase above 8 T}

\begin{figure}
\includegraphics[width=1.0\linewidth]{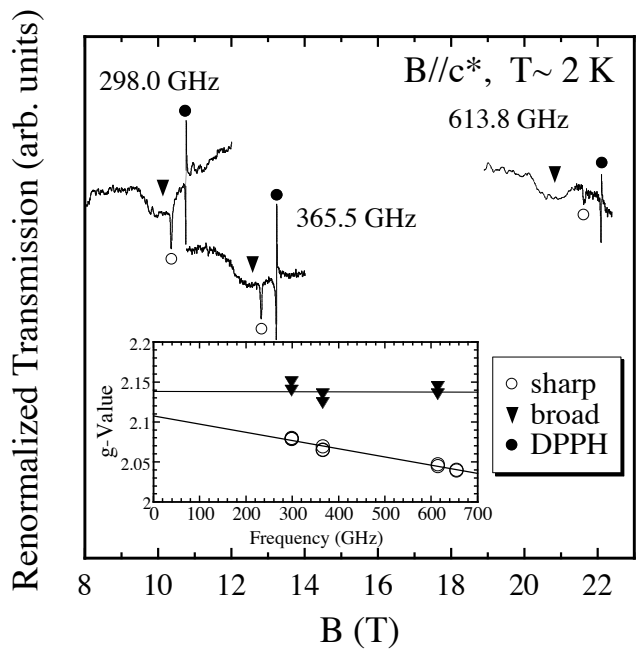}%
\caption{\label{fig:five} ESR spectra of
$\lambda$-(BETS)$_2$Fe$_{0.6}$Ga$_{0.4}$Cl$_4$ at high magnetic
fields. The inset shows the frequency dependence of the effective \textit{g}-value
for the sharp (open circle) and broad (triangle) absorption lines.
The sharp absorption
represented by the solid circle is DPPH, used as a field marker.
}
\end{figure}

To explore the high field ground states, we employed the single-pass transmission 
spectroscopy technique in combination with the 25 T resistive magnet, 
which covers the PM phase and the FISC
phase. Typical spectra at $\sim$2 K with the magnetic field
applied parallel to the \textit{c}$^*$-axis are shown in Fig.
\ref{fig:five}. Each spectrum is renormalized since the power of
the light source depends on the observing frequency. Two
absorption lines are observed, one with a broad linewidth and the
other with a sharp linewidth, represented by triangles and open
circles in Fig. \ref{fig:five}, respectively. 
We note that the frequency 298.0 GHz corresponds to $\sim$10 T
when \textit{g}=2 which corresponds to the boundary of the FISC
phase (see Fig. \ref{fig:one}). Therefore, for higher frequencies,
the ESR at low temperatures is measured in the FISC phase. As
shown in Fig. \ref{fig:five}, although the sharp absorption line
intensity and linewidth become weaker and broader, the broad
absorption lines do not change as the observing frequency
increases. Moreover, the effective \textit{g}-value of the sharp absorption
depends on frequency, while the broad one does not (see the inset
of Fig. \ref{fig:five}). The details of these absorption lines
will be discussed in the next section.

\begin{figure}
\includegraphics[width=0.95\linewidth]{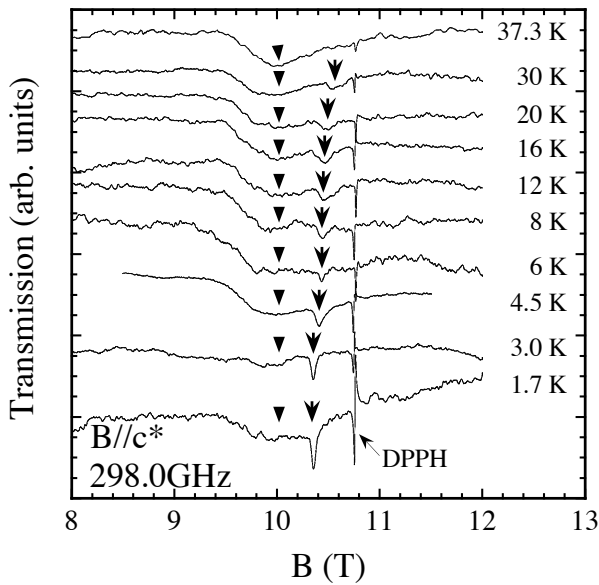}%
\caption{\label{fig:six} The temperature dependence of the typical spectra for $\lambda$-(BETS)$_2$Fe$_{0.6}$Ga$_{0.4}$Cl$_4$
at high magnetic field. The sharp and broad absorption lines are represented by the arrow and triangle, respectively. }
\end{figure}

Figure \ref{fig:six} shows the temperature dependence of ESR
spectra for 298.0 GHz for \textit{B}//\textit{c}$^*$-axis.
Although the broad absorption line is almost independent of
temperature, the sharp absorption line appears around 30 K and
shifts gradually as the temperature decreases. The same behavior
is also observed in the higher frequency data. The sharp
absorption intensity increases as the temperature is decreased,
which means the sharp absorption is EPR. 

\section{Discussion}

In this section, we discuss the nature of the ground state of
$\lambda$-(BETS)$_2$Fe$_{0.6}$Ga$_{0.4}$Cl$_4$ and the specific
details of its phase diagram, as probed by the ESR study.

\subsection{AFI phase}

In the AFI phase, we have observed AFMR which can be described by the biaxial AFMR theory
and we have obtained two parameters, $\sqrt{C_1}$= $B_{sf}$ =0.9 T and $\sqrt{C_2}$=1.1 T.
The spin-flop field $B_{sf}$ is consistent with the magnetic susceptibility results of
$\lambda$-(BETS)$_2$Fe$_x$Ga$_{1-x}$Cl$_4$ where $B_{sf}$ is 0.7 T and 0.75 T
for \textit{x}=0.55 and \textit{x}=0.7, respectively.\cite{alloy-akane}
The exchange field $H_E$ can be given by $H_E=AM_0$ where \textit{A} is the molecular field coefficient
and $M_0$ is the magnitude of the sublattice moment in the antiferromagnetic state.
These values can be obtained from the following relationships,
\begin{eqnarray}
A\sim 1/\chi_{\mathit{peak}},
\label{eq:molecularField}
\end{eqnarray}
\begin{eqnarray}
M_0=\frac{N}{2} g \mu{_B} S,
\label{eq:magneticMoment}
\end{eqnarray}
where $\chi_{\mathit{peak}}$ is the magnitude of the magnetic
susceptibility peak and \textit{N} is number of spins. The
susceptibility of $\lambda$-(BETS)$_2$Fe$_x$Ga$_{1-x}$Cl$_4$ at
$T_N$ is around 0.22 emu/mol, which is relatively high for organic
conductors.\cite{alloy-akane, alloy-akutu} This is due to the
contribution of the magnetic Fe$^{3+}$ moments. If we assume that
\textit{g} and \textit{S} are 2 and 5/2, respectively, the
obtained exchange field is $H_E \sim$6.3 T which is similar value
to the pure FeCl$_4$ salts.\cite{brossard} The relation between
the exchange field $H_E$ and the exchange interaction \textit{J}
can be expressed by
\begin{eqnarray}
H_E=\frac{2 z J S}{g \mu{_B}},
\label{eq:exchangeField}
\end{eqnarray}
where \textit{z} is the number of the nearest-neighbor spins. If
we assume that the nearest-neighbor for Fe$^{3+}$ moments is 2
along the \textit{a}-axis,\cite{brossard, crystal} then the value
of \textit{J} is obtained to be 0.8 K, which is in a good
agreement with the exchange interaction $J_d$ calculated from
mean-field theory.\cite{mori} If we use $H_E \sim$6.3 T, the
anisotropy field $H^1_A$ and $H^2_A$ can be obtained from $C_1$
and $C_2$ , respectively, as mentioned in Sec. 3. Then we obtain
$H^1_A$=38 mT, $H^2_A$ =64 mT. The exchange field is found to be
much larger than the anisotropy fields, $H^i_A$ (\textit{i}=1,2)
$\ll H_E$, which is consistent with the appearance of the
spin-flop transition.\cite{alloy-akane} The anisotropy field ratio of
$H^2_A / H^1_A$ is about 2. We note that the values of $H_E$,
$H^1_A$, and $H^2_A$ for pure FeCl$_4$ salts obtained by Suzuki
\textit{et al}. are 1-2 orders different from our results.\cite{toyota} 
This difference arises from the assumption that the
AFMR comes from the spin wave excited in the $\pi$ spin system or
\textit{d} spin system. Although the strong interaction between 
$\pi$-$\pi$ and $\pi$-\textit{d} electrons can not be ignored, 
it is difficult to explain the high magnetic susceptibility of this material at
$T_N$ if the exchange field is 2 orders higher than our result.

In principle, the intensity of EPR just above $T_N$ is comparable
to the intensity of AFMR as observed in Fig. \ref{fig:two} (b).
Since the EPR is likely to originate from Fe$^{3+}$ due to the
large magnetic moment, the AFMR should also involve the
\textit{d}-electrons. It is also important to note that usually
the resonances from $\pi$- and \textit{d}-electrons merge into one
resonance due to the exchange interaction between them. The
resonances can be split when the Zeeman energy exceeds the
exchange interaction, $2J \sim \Delta g \mu{}_B H$ where $\Delta
g$ is the difference of the \textit{g}-values between the two
spins.\cite{exchangenarrow} The exchange interaction
\textit{J}$_{\pi{}-d}$ is estimated to be $\sim$15 K for FeCl$_4$
salts.\cite{mori} Then the Zeeman energy is around 625 GHz which
is much higher than the observing frequency. Therefore, the
resonances should combine as one resonance which is predominantly 
from the S=5/2 \textit{d}-electrons, especially at
low temperature.

We have found that the easy-axis is near the \textit{b}-axis, based on our fit to the
AFMR theory as shown in Fig. \ref{fig:four}. This is distinct from
the results for the pure FeCl$_4$ salt where the easy-axis is
found to be in a direction tilted by 30$^\circ$ from \textit{c}-
to \textit{b}$^*$-axes.\cite{toyota, torque1, torque2} The
Fe$^{3+}$ content of the alloy we have investigated is 40 \% less
than the pure FeCl$_4$ salt. This implies that the distances
between \textit{d}-electrons are, on average, larger, and this may
be the reason for the easy-axis difference. 
In fact, from studies of the spin-Peierls compound CuGeO$_3$,\cite{cugeo3} 
it is well-known that doping can significantly affect the magnetic
anisotropy (Ref. 18 and Refs. therein). The easy-axis is
probably not exactly along the \textit{b}-axis since there is some
uncertainty in the theoretical fits in Fig. \ref{fig:four},
particularly at low frequencies. Angular dependent torque
measurements would be useful to determine the exact direction of
the easy-axis, and to follow the change of easy-axis direction
systematically with the Fe$^{3+}$ concentration.

\subsection{PM and FISC phases}

We now discuss the EPR observed in the high field region.  Two
absorption lines have been observed where the broad signal is
independent of temperature, and the sharp signal is temperature
and frequency dependent. Here the magnetic field is aligned along
the \textit{c}$^*$-axis. Surprisingly, the two EPR absorption
lines are still observed at 2 K in the FISC state, as shown in
Figs. \ref{fig:one} and \ref{fig:five}. Normally, the ESR
measurement in a superconductor becomes difficult due to the
limitation of the penetration depth $\lambda$. Although the
penetration depth for this system is unknown, there should be some
penetration of the magnetic flux if we consider that the
Jaccarino-Peter compensation effect takes place. In this case, the
magnetic field should be inhomogeneous, which might cause linewidth
broadening or a \textit{g}-shift of the observed absorption. As
mentioned above, the sharp absorption becomes broader and its
\textit{g}-value is changing as the frequency increases at $\sim$2
K (see Fig. \ref{fig:five}). The intensity of the sharp signal
also decreases. However, the temperature dependence of the
linewidth and the \textit{g}-value do not change significantly at
the phase boundary, as shown in Fig. \ref{fig:six}. This suggests
that the behavior of the sharp resonance (i.e. linewidth
broadening and \textit{g}-shift at $\sim$2 K) is not related to
the FISC state. We think that the sample is only partially
superconducting at $\sim$2 K and the resonance is coming from
paramagnetic domains.

\begin{figure}
\includegraphics[width=0.95\linewidth]{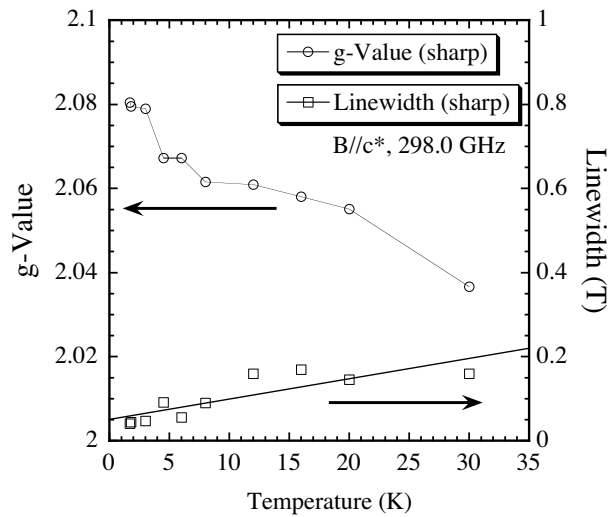}%
\caption{\label{fig:seven} The temperature dependence of the effective \textit{g}-value and the linewidth taken
from the sharp absorption observed in Fig. 5.}
\end{figure}

The shift of EPR line versus the temperature,
which is known as the \textit{g}-shift, usually occurs as a result
of the spin-orbit interaction. However, in the case of magnetic
metals, the interaction between the conduction electron and the
localized spins (i.e. $\pi$-\textit{d} interaction) also causes a
\textit{g}-shift. Therefore, it is possible that the
$\pi$-\textit{d} interaction \textit{J}$_{\pi{}-d}$ 
increases at lower temperature, leading to the
\textit{g}-shift of the EPR line.
According to the basic ESR theory of magnetic ions in metals,
the temperature dependence of the
linewidth and the \textit{g}-shift can be expressed as
\begin{eqnarray}
\Delta{} H=A+BT,
\label{eq:sdLinewidth}
\end{eqnarray}
\begin{eqnarray}
\Delta{} g=J_{\pi{}-d}D(E_F),
\label{eq:sdGvalue}
\end{eqnarray}
respectively.\cite{s-desr} Here, $D(E_F)$ is the density of state
at the Fermi level, \textit{A} is a residual linewidth, and
\textit{B} is given by
\begin{eqnarray}
B=\frac{k_B \pi (\Delta{}g)^2}{g_{\pi} \mu{_B}},
\label{eq:korringaRelation}
\end{eqnarray}
known as the ``Korringa relation". Figure \ref{fig:seven} shows
the temperature dependence of the effective \textit{g}-value and linewidth
for the sharp absorption. A linear fit to the linewidth yields
\textit{A}=0.05 T and \textit{B}=4.8$\times$10$^{-3}$ T/K. If we
assume the \textit{g}-value of the $\pi$-electrons is $g_{\pi}$=2
and use the parameter \textit{B} in Eq. \ref{eq:korringaRelation},
we obtain $\Delta{}g$=0.05 which is in good agreement with our
results of Fig. \ref{fig:seven}. However, this is not the case if
we derive the $\pi$-\textit{d} exchange interaction from Eq.
\ref{eq:sdGvalue}. If we estimate the density of states at Fermi
level $D(E_F)$ to be 1.2$\times$10$^{22}$ J$^{-1}$ from the
specific heat measurements,\cite{hinetu} then
\textit{J}$_{\pi{}-d}$ becomes 8 mK, which is not consistent with
the theoretical value.\cite{mori} This discrepancy comes from the
so-called ``bottleneck effect" where \textit{J}$_{\pi{}-d}$ is
often underestimated.\cite{s-desr} In the present case, we assume
an effective \textit{g}-value of combined $\pi$- and
\textit{d}-electron spins. Although the intensity of the
conduction electron spin resonance is temperature independent in
response to the Pauli spin susceptibility, the spin susceptibility
of localized \textit{d}-electron increases with decreasing
temperature. Therefore, the effective \textit{g}-value usually
passes from that for conduction electrons at high temperatures to
that for localized spins at lower temperatures.\cite{s-desr} The
behavior in Fig. \ref{fig:six} is consistent with this prediction
(i.e. the \textit{g}-value is close to 2 at high temperature) and
suggests the presence of non-negligible $\pi$-\textit{d}
interactions in the system.

\begin{figure}
\includegraphics[width=0.95\linewidth]{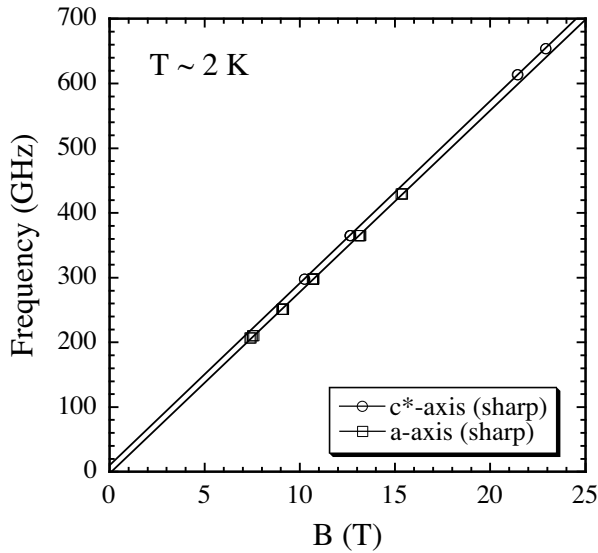}%
\caption{\label{fig:eight} The frequency-field diagram of the observed EPR (sharp absorption) 
for \textit{B}//\textit{c}$^*$- and \textit{a}-axes.}
\end{figure}

The frequency dependence of the effective \textit{g}-value for the sharp
absorption line is also of note (see inset of Fig.
\ref{fig:five}). To explain this behavior, we have considered a
system where the spins of $\pi$- and \textit{d}-electrons,
$S_{\pi}$=1/2 and $S_d$=5/2, make a spin pair. In this case, the
Pryce's spin Hamiltonian can be expressed as follows,
\begin{eqnarray}
H=g \mu{_B} H_0
\left( S_{dz} + S_{\pi{}z} \right)
+D S_{dz}^2
-J \left( \mathbf{S_d} \cdot \mathbf{S_{\pi}} \right).
\label{eq:spinHamiltonian}
\end{eqnarray}
Here, the first term is the Zeeman part where the two spins
combine to form an effective \textit{g}-value. The second term is
the anisotropic \textit{d}-electron spin term. This term is
necessary since the Cl atoms in the FeCl$_4$$^-$ anion are very
close to the donor BETS molecules,  leading to lower symmetry in
the ligand field.\cite{crystal} The last term is the exchange
interaction term where only the interaction between $\pi$- and
\textit{d}-electrons is taken into account for simplicity. If we
assume $J \gg D$ and $g \mu{}_B H_0$, the combined spins
\textit{S}=3, 2 and the \textit{z} components become good quantum
numbers and we obtain 12 energy states. If we consider only the
low energy state, the \textit{S}=2 energy states are lower than
the \textit{S}=3 states when $J<0$. Then, we have
\begin{eqnarray}
E_2=2 g \mu{_B} H_0 + \left( \frac{17}{4} \right) D+ \left( \frac{7}{2} \right) J,\nonumber \\
E_1=g \mu{_B} H_0 + \left( \frac{5}{4} \right) D+ \left( \frac{7}{2} \right) J, \nonumber \\
E_0=\left( \frac{1}{4} \right) D+ \left( \frac{7}{2} \right) J, \\
E_{-1}=-g \mu{_B} H_0 + \left( \frac{5}{4} \right) D+ \left( \frac{7}{2} \right) J, \nonumber \\
E_{-2}=-2 g \mu{_B} H_0 + \left( \frac{17}{4} \right) D+ \left( \frac{7}{2} \right) J. \nonumber
\label{eq:energyLevels}
\end{eqnarray}
Therefore, the allowed transitions $\Delta{}M=\pm 1$ in ESR are
\begin{eqnarray}
h \nu{}= g \mu{_B} H_0 + 3 D, \nonumber \\
h \nu{}= g \mu{_B} H_0 + D, \\
h \nu{}= g \mu{_B} H_0 - D, \nonumber \\
h \nu{}= g \mu{_B} H_0 - 3 D. \nonumber
\label{eq:resonanceCondition}
\end{eqnarray}
The excitation from the lowest energy state is from $E_{-2}$ to
$E_{-1}$ in this model, and the ESR at low temperature should
mainly involve this transition, i.e. $h \nu{}= g \mu{_B} H_0 - 3
D$. We show the frequency-resonance field diagram of the sharp
absorption in Fig. \ref{fig:eight}. The resonance plots for
\textit{B}//\textit{c}$^*$ (open circles) and the linear fit show
that there is an offset of 9.9 GHz from the origin, which is
consistent with the present model. Therefore, if we take $D<0$, we
obtain $|D|\sim$0.11 cm$^{-1} \sim$ 0.16 K and
\textit{g}$\sim$2.01.

The anisotropic \textit{D} term varies with the orientation of the
magnetic field. Thus, if we apply the field to the
\textit{a}-axis, the sharp absorption shifts to higher field as
shown in Fig. \ref{fig:eight} (rectangles). We note that this
model is not limited to the
$\lambda$-(BETS)$_2$Fe$_{0.6}$Ga$_{0.4}$Cl$_4$ salts,  but 
this kind of resonance should be seen in any system where there is
a non-negligible $\pi$-\textit{d} interaction. Indeed, the same
kind of absorption behavior is observed for (DMET)$_2$FeBr$_4$
above the saturation field.\cite{okubo} In our model, the
anisotropic parameter for (DMET)$_2$FeBr$_4$ would be $|D|\sim$0.3
cm$^{-1}$ for \textit{B}//\textit{c}$^*$, which is similar to our
findings for $\lambda$-(BETS)$_2$Fe$_{0.6}$Ga$_{0.4}$Cl$_4$.

Finally, we address the origin of the broad absorption line,
which, unlike the sharp absorption line, does not show any
significant change in \textit{g}-value or linewidth. Moreover, the
absorption does not shift with axis orientation. At present we
have no explanation for the origin of this resonance, but it could
be a result of impurities or disorder in the alloy system.

\section{Summary}

In summary, we have performed ESR measurements on the organic 
$\pi$-\textit{d} electron system
$\lambda$-(BETS)$_2$Fe$_{0.6}$Ga$_{0.4}$Cl$_4$. The alloy system
was selected so that we could carry out ESR experiments in all low
temperature, field dependent ground states, including the field
induced superconducting (FISC) state. We have observed AFMR and
EPR for the AFI phase and PM phase, respectively. The
antiferromagnetic order shows a biaxial anisotropic behavior which
originates from the exchange interaction between
\textit{d}-electrons. The easy and hard axes correspond to the
\textit{b}- and \textit{c}$^*$-axes, respectively, and the
spin-flop field is around $\sim$0.9 T. The EPR shows behavior that
is related to significant $\pi$-\textit{d} interaction and
anisotropic Fe$^{3+}$ magnetic moments. Due to the bottleneck
effect, we could not estimate directly the exchange interaction
between $\pi$- and \textit{d}-electrons \textit{J}$_{\pi{}-d}$.
Usually, the addition of non-magnetic impurities contributes to the 
breaking of the bottleneck.\cite{s-desr} Hence, the ESR
measurements of the $\lambda$-(BETS)$_2$Fe$_x$Ga$_{1-x}$Cl$_4$
alloys with a small portion of \textit{x} might be promising to
estimate \textit{J}$_{\pi{}-d}$. Finally, we did not observe any
significant change in the EPR signal at high fields between the
normal and FISC ground states. This may be a result of the nature
of the FISC state, which, in the Jaccarino-Peter scenario, would
allow flux to penetrate the bulk sample.

\begin{acknowledgments}
Y. O. acknowledges Dr. S. Uji and Prof. H. Ohta for helpful discussions.
This work has been funded by NHMFL/IHRP 5042 and NSF-DMR-0203532.
The NHMFL is supported by a contractual agreement between NSF and the state of Florida.
\end{acknowledgments}



\begin{thebibliography}{99}
\bibitem{ujinature}
S. Uji, H. Shinagawa, T. Terashima, T. Yakabe, Y. Terai, M. Tokumoto, A. Kobayashi, H. Tanaka and H. Kobayashi, Nature 410 (2001) 908.
\bibitem{j-peffect}
V. Jaccarino and M. Peter, Phys. Rev. Lett. 9 (1962) 290.
\bibitem{alloy-uji}
S. Uji, T. Terashima, C. Terakura, T. Yakabe, Y. Terai, S. Yasuzuka, Y. Imanaka, M. Tokumoto, A. Kobayashi, F. Sakai, H. Tanaka, H. Kobayashi, L Balicas and J. S. Brooks, J. Phys. Soc. Jpn. 72 (2003) 369.
\bibitem{luispr}
L. Balicas,  J. S. Brooks, K. Storr, S. Uji, M. Tokumoto, H. Tanaka, H. Kobayashi, A. Kobayashi, V. Barzykin and L. P. Gorkov, Phys. Rev. Lett. 87 (2001) 067002.
\bibitem{brossard}
L. Brossard, R. Clerac, C. Coulon, M. Tokumoto, T. Ziman, D. K. Petrov, V. N. Laukhin, M. J. Naughton, A. Audouard, F. Goze, A. Kobayashi, H. Kobayashi and P. Cassoux, Eur. Phys. J. B 1 (1998) 439
\bibitem{toyota}
T. Suzuki, H. Matsui, H. Tsuchiya, E. Negishi, K. Koyama and N. Toyata, Phys. Rev. B, 67 (2003) 020408(R).
\bibitem{isaak}
I. Rutel, S. Okubo, J. S. Brooks, E. Jobiliong, H. Kobayashi, A. Kobayashi and H. Tanaka, Phys. Rev. B 68 (2003) 144435.
\bibitem{crystal}
H. Kobayashi, H. Tomita, T. Naito, A. Kobayashi, F. Sakai, T. Watanabe and P. Cassoux, J. Am. Chem. Soc. 118(1996) 368.
\bibitem{mvna}
S. Hill, N. S. Dalal and J. S. Brooks, Appl. Magn. Reson. 16 (1999)237.
\bibitem{sergei}
S. A. Zvyagin, J. Krzystek, P. H. M. van Loosdrecht, G. Dhalenne and A. Revcolevschi, Physica B 346-347 (2004) 1.
\bibitem{cugeo3}
K. Uchinokura, J. Phys.: Condens. Matter 14 (2002) R195.
\bibitem{afmr}
T. Nagamiya, K. Yoshida and R. Kubo, Adv. Phys. 4 (1955) 1.
\bibitem{alloy-akane}
A. Sato, E. Ojima, H. Akutsu, Y. Nakazawa, H. Kobayashi, H. Tanaka, A. Kobayashi and P. Cassoux, Phys. Rev. B 61 (2000) 111.
\bibitem{alloy-akutu}
H. Akutsu, K. Kato, E. Ojima, H. Kobayashi, H. Tanaka, A. Kobayashi and P. Cassoux, Phys. Rev. B 58 (1998) 9294.
\bibitem{mori}
T. Mori and M. Katsuhara, J. Phys. Soc. Jpn. 71 (2002) 826.
\bibitem{exchangenarrow}
P. W. Anderson, J. Phys. Soc. Jpn. 9 (1954) 316.
\bibitem{torque1}
J. I. Oh, M. J. Naughton, T. Courcet, I. Malfant, P. Cassoux, M. Tokumoto, H. Akutsu, H. Kobayashi and A. Kobayashi, Synth. Met. 103 (1999) 1861.
\bibitem{torque2}
T. Sasaki, H. Uozaki, S. Endo and N. Toyota, Synth. Met.,120 (2001) 759.
\bibitem{s-desr}
R. H. Taylor, Adv. Phys. 24 (1975) 681.
\bibitem{hinetu}
E. Negishi, H. Uozaki, Y. Ishizaki, H. Tsuchiya, S. Endo, Y. Abe, H. Matsui and N. Toyota, Synth. Met. 133-134 (2003) 555.
\bibitem{okubo}
S. Okubo, K. Kirita, Y. Inagaki, H. Ohta, K. Enomoto, A. Miyazaki and T. Enoki, Synth. Met. 135-136 (2003) 589.
\end {thebibliography}

\end{document}